\documentclass[conference]{IEEEtran}
\IEEEoverridecommandlockouts
\usepackage[style=ieee,url=false,doi=false,isbn=false]{biblatex}

\usepackage{amsmath,amssymb,amsfonts}
\usepackage{algorithm}
\usepackage{algpseudocode}
\usepackage{graphicx}
\usepackage{pdfpages}
\usepackage{textcomp}
\usepackage{xcolor}
\usepackage{braket}
\usepackage{physics}
\usepackage{subfigure}
\usepackage{bm}
\usepackage{subcaption}
\usepackage{subfig}
\usepackage{enumitem}

\DeclareRobustCommand{\erase}{\bgroup\markoverwith{\textcolor{red}{\rule[.5ex]{2pt}{0.4pt}}}\ULon}

\def\BibTeX{{\rm B\kern-.05em{\sc i\kern-.025em b}\kern-.08em
    T\kern-.1667em\lower.7ex\hbox{E}\kern-.125emX}}

\begin{document}

\title{
  Enhancing NDAR with Delay-Gate-Induced Amplitude Damping
  \thanks{
    This work was performed for Council for Science, Technology and Innovation (CSTI), Cross-ministerial Strategic Innovation Promotion Program (SIP), ``Promoting the application of advanced quantum technology platforms to social issues'' (Funding agency: QST).
  }
}

\makeatletter
\newcommand{\linebreakand}{%
\end{@IEEEauthorhalign}
\hfill\mbox{}\par
\mbox{}\hfill\begin{@IEEEauthorhalign}
}
\makeatother

\author{
  \IEEEauthorblockN{
    Wai-Hong Tam,
    Hiromichi Matsuyama,
    Ryo Sakai,
    and
    Yu Yamashiro
  }
  \IEEEauthorblockA{
    \textit{Jij Inc.}, 3-3-6 Shibaura, Minato-ku, Tokyo, 108-0023, Japan
  }

  \IEEEauthorblockA{
    w.tam@j-ij.com,
    h.matsuyama@j-ij.com,
    r.sakai@j-ij.com,
    y.yamashiro@j-ij.com,
  }
}

\maketitle

\begin{abstract}
  The Noise-Directed Adaptive Remapping (NDAR) method utilizes amplitude damping noise to enhance the performance of quantum optimization algorithms.
  The NDAR method alternates between exploration by sampling solutions from the quantum circuit and exploitation by transforming the cost Hamiltonian by changing the signs of its terms.
  Both exploration and exploitation are important components in the design of classical heuristic algorithms.
  In this study, we examine how NDAR performance is improved by adjusting the balance between these two components.
  We control the degree of exploitation by varying the delay time to 0, 50, and 100$\mu$s and investigate different exploration strategies using two quantum circuits, the Quantum Approximate Optimization Algorithm (QAOA) and a random circuit, on IBM’s Heron processor.
  Our results show that increasing the delay time in the NDAR method improves the best objective value found in each NDAR iteration.
  In the cases of single-layer QAOA circuits and random circuits, sampling applied to unweighted-sparse MaxCut problems with low edge density, both exploration strategies yield similar objective value trajectories and provide competitive solution quality to the simulated annealing method for the 80-node problem.
  These findings demonstrate the effectiveness of NDAR, while the similar performance of both exploration strategies indicates that, in most cases, increasing amplitude damping noise through additional delay time results in information loss within quantum circuits.
  On the other hand, QAOA outperforms random circuits in some specific cases, like positive-negative weighted MaxCut problems on the fully connected graph.
  This observation indicates the potential advantages of QAOA in more complex problem settings.
  Furthermore, we develop the classical NDAR to acquire a deeper understanding of the effects of different exploration strategies.
  We demonstrated that controlling the Hamming weight distribution of sampled bitstrings provides higher quality solutions.
  This classical result suggests that if we find an appropriate quantum circuit for exploration, the performance of the NDAR method could be enhanced.
\end{abstract}


\section{Introduction}
\label{sec:intro}

Combinatorial optimization is an interdisciplinary research area in computer science, mathematics, and statistical physics, with applications in logistics, finance, and machine learning~\cite{papadimitriou1998, schrijver2003}.
Many combinatorial optimization problems, such as the MaxCut, graph partitioning, and constraint satisfaction problems, are computationally challenging due to their NP-hard nature~\cite{karp1972, garey1979}.
To tackle these problems, researchers have developed not only exact algorithms that guarantee optimality but also numerous metaheuristic algorithms that provide high-quality solutions within reasonable computational time~\cite{Gendreau2018-jm, Blum2003, Fister2013}.
Metaheuristic algorithms are powerful tools for delivering practical solutions of sufficient quality to complex real-world problems while requiring only feasible computational resources.

Quantum computing has emerged as a promising approach to combinatorial optimization~\cite{montanaro2016}.
The Quantum Approximate Optimization Algorithm (QAOA)~\cite{farhi2014, Blekos:2023nil} represents a heuristic approach to solving combinatorial optimization problems using Noisy Intermediate Scale Quantum (NISQ) devices~\cite{Abbas:2023agz}.
This algorithm hybridizes quantum and classical computing by employing a quantum circuit with adjustable parameters to evaluate the cost Hamiltonian.
A classical optimizer then tunes these parameters to achieve the minimum expectation value of the cost Hamiltonian.
The final solution to the optimization problem corresponds to the ground state of the cost Hamiltonian.

However, noise within quantum devices presents significant challenges in quantum computing, particularly by limiting the circuit depth and computational effectiveness.
In the NISQ era, error mitigation is a common strategy to address noise in quantum devices.
Error mitigation techniques~\cite{Temme2017, Endo2021} improve output quality from quantum devices, enhancing both expectation values~\cite{Temme2017} and sampling distributions~\cite{Nation2021}, though with increased sampling and classical computation costs.
While these techniques typically treat noise as a detrimental factor, an alternative perspective explores how certain quantum noise types can be leveraged to enhance optimization.
The Noise-Directed Adaptive Remapping (NDAR) method~\cite{Filip2024} strategically harnesses quantum noise rather than mitigating it.
In noisy quantum devices, certain computational basis states---termed noise attractor states---appear more frequently due to phenomena such as amplitude damping noise.
The NDAR method modifies the cost Hamiltonian to assign favorable solutions to these noise attractor states.
It then iteratively searches for solutions by alternating between cost Hamiltonian modifications and sampling under the updated Hamiltonian.
This approach has been explored in several studies~\cite{Filip2024, Maciejewski2024}, demonstrating its potential applicability on noisy devices.

To design practical algorithms, efficiently searching large and complex solution spaces remains crucial.
In classical heuristic algorithms, finding an appropriate balance between exploration---searching for new, potentially better solutions---and exploitation---leveraging known high-quality solutions to search for better solutions---is essential for designing practical optimization approaches.
This trade-off is known as the exploration-exploitation dilemma~\cite{Sutton1998} in numerous fields, including optimization, reinforcement learning, and evolutionary computation~\cite{Eiben1998}.
The NDAR method can also be conceptualized as an algorithm that balances exploration through quantum circuit sampling and exploitation through cost Hamiltonian modification.
In this study, we investigate this balance in the NDAR method and its impact on performance.
To control the degree of exploitation, we introduce delay gates before measurement and systematically vary their durations.
For exploration, we examine different quantum circuits for sampling: QAOA and random circuits.

We conducted experiments on IBM's Heron quantum processor to assess the effectiveness of the NDAR method.
We evaluated its performance on MaxCut problems on $80$ node unweighted-sparse and weighted-dense graphs.
We compared two exploration strategies, QAOA and random circuits, under amplitude damping noise whose strength is tuned by delay time $T_{\mathrm{d}}$.
In the low-density unweighted MaxCut problem, both circuits achieved strong performance under $T_{\mathrm{d}} = 50~\mu$s and $100~\mu$s, with final $E_{\mathrm{Best}} / E_{\mathrm{SA}}$ values exceeding 0.9,
where $E_{\mathrm{Best}}$ and $E_{\mathrm{SA}}$ represent the best energy found by NDAR method and a simulated annealing method, respectively.
For the weighted-dense case, similar performance trends were observed across different delay times, as in the unweighted-sparse case.
However, the final solution quality was generally lower.
Notably, at a delay time of $T_{\mathrm{d}} = 50~\mu$s, the QAOA circuit achieved slightly better performance than the random circuit.

While quantum implementations on current hardware are limited to systems of around 100 qubits, we also propose a classical variant of NDAR that replaces quantum sampling with a simple probabilistic sampling scheme, where each bit is set to 0 with probability $q$.
This not only addresses the scalability constraint but also serves to make our heuristic algorithm perspective more concrete.
This classical NDAR emulates random-circuit NDAR and serves as a baseline for performance evaluation.
We conducted experiments on $80$-node and $300$-node MaxCut problems.
In the $80$-node case, the classical NDAR exhibits behavior similar to the quantum NDAR.
The classical emulation result suggests that the quantum NDAR would work effectively even for larger problem instances.

This paper is organized as follows.
In Sec.~\ref{sec:method}, we introduce QAOA and the NDAR method.
We present our proposed method for controlling the exploration and exploitation strategies in Sec.~\ref{sec:proposed_method}.
Section~\ref{sec:result} provides experimental results under different exploration and exploitation settings across the problems.
We developed the classical NDAR method inspired by the experimental results on the real device in Sec.~\ref{sec:classical}.
Finally, in Sec.~\ref{sec:summary}, we summarize the work and discuss directions for future research.

\section{Background}
\label{sec:method}

\subsection{Quantum Approximate Optimization Algorithm}
In combinatorial optimization, the goal is to minimize a given cost function \( C(\bm{x}) \) that takes an $n$-bit string \( \bm{x} = x_1x_2 \cdots x_n\) with $x_i \in \{0,1\}$ as an argument.
One can typically encode such an optimization problem into the Ising model~\cite{Lenz1920, Ising1925}:
\begin{equation}
  C(\bm{s}) = \sum_i h_i s_i + \sum_{i<j} J_{ij} s_i s_j,
  \label{eq:isingcost}
\end{equation}
where \( s_i = \pm1 \) represents classical spin variables, and $h_i$ and $J_{ij}$ represent the strength of the external field and the interactions, respectively.
Here, the relationship between a binary variable $x_i$ and a spin variable $s_i$ can be written as $s_i = 1 - 2x_i$.
By mapping spin variables $s_i$ to the eigenvalues of the Pauli $Z$ operator in the computational basis, we can construct the cost Hamiltonian
\begin{equation}
  H_C= \sum_i h_i Z_i + \sum_{i<j} J_{ij} Z_i Z_j,
  \label{eq:cost_function}
\end{equation}
where $Z_i$ represents the Pauli $Z$ operator acting on the $i$-th qubit.
With this definition, the expectation value of cost Hamiltonian \( H_C \) in the computational basis corresponds to the cost function:
\begin{equation}
  \Braket{\bm{x} | H_C | \bm{x}} = C(\bm{x}).
\end{equation}
Thus, solving the optimization problem is equivalent to finding the ground state of \( H_C \).

One promising approach for approximating such ground states on quantum hardware is the Quantum Approximate Optimization Algorithm (QAOA)~\cite{farhi2014}.
A QAOA circuit consists of multiple layers, each applying two unitary transformations: Cost Hamiltonian evolution, \( {U_P(\gamma_l) = } \exp(-i\gamma_l H_C) \), and Mixer Hamiltonian evolution, \({U_M(\beta_l) = } \exp(-i\beta_l H_{\text{mix}}) \), where \( l \) denotes the layer index.
The mixer Hamiltonian \( H_{\text{mix}} \) is typically chosen as
\begin{equation}
  H_{\text{mix}} = \sum_j X_j,
\end{equation}
ensuring that every qubit can transition independently between computational basis states.
Applying the alternating unitaries for \( p \) layers yields the final QAOA state:
\begin{equation}
  \Ket{\psi_p(\boldsymbol{\gamma}, \boldsymbol{\beta})} = \prod_{l=1}^{p} e^{-i \beta_l H_{\text{mix}}} e^{-i \gamma_l H_C} |+\rangle^{\otimes n},
  \label{eq:qaoa_state}
\end{equation}
where $\ket{+}^{\otimes n} = (1/\sqrt{2^n}) \sum_{\bm{x}} \ket{\bm{x}}$ is the uniform superposition state.
The expectation value of the cost Hamiltonian is measured as:
\begin{equation}
  \Braket{\psi_p(\boldsymbol{\gamma}, \boldsymbol{\beta}) | H_C | \psi_p(\boldsymbol{\gamma}, \boldsymbol{\beta})}.
\end{equation}
To minimize this expectation value, the variational parameters \( \boldsymbol{\gamma} = (\gamma_1, \gamma_2, \dots, \gamma_p) \) and \( \boldsymbol{\beta} = (\beta_1, \beta_2, \dots, \beta_p) \) are optimized iteratively on a classical device.
Since each layer introduces two independent parameters, a \( p \)-layer QAOA circuit has \( 2p \) tunable parameters.

Increasing the number of layers \( p \) increases the expressibility of the ansatz.
However, higher values of \( p \) also introduce several challenges.
The first challenge involves the increasing computational cost of optimizing variational parameters due to the growth in the number of parameters in the variational circuit.
The second challenge is the amplified noise due to the growing number of multi-qubit gates in the cost unitary operator and the decoherence resulting from longer execution times.
In later sections, we will explain how the NDAR method deals with noises.

\subsection{Noise-Directed Adaptive Remapping}
Quantum optimization algorithms, such as QAOA, often suffer from performance degradation due to noise, which disrupts the quantum states.
On NISQ devices, quantum error mitigation techniques represent an important approach for reducing the impact of noise on calculation results.
These techniques require overhead for both classical and quantum computational costs.
The NDAR method~\cite{Filip2024} takes an alternative approach by embracing noise as an advantage rather than a flaw.
In noisy quantum systems, certain states tend to emerge more frequently due to noise.
We refer to these states as noise attractor states.
NDAR applies iterative bit-flip gauge transformations to the cost Hamiltonian, aligning these noise attractor states with high-quality solutions.
With this technique, the NDAR method transforms the detrimental effects of noise into a useful bias that steers the optimization process toward better outcomes.

First, we consider the unitary bit-flip operator acting on all qubits, specified by the bitstring \( \boldsymbol{y} \in \{0,1\}^n \), given by
\begin{equation}
  P_{\boldsymbol{y}} = \bigotimes_{i=0}^{n-1} X_i^{y_i}.
\end{equation}
Applied to $\ket{0\cdots 0}$, this bit-flip operator provides
\begin{equation}
  P_{\boldsymbol{y}} \Ket{0\cdots 0}=  \Ket{y_{n-1} \cdots y_0} .
\end{equation}
Next, we consider the result of applying $P_{\boldsymbol{y}}$ to $H_C$ in Eq.~\eqref{eq:cost_function}.
The cost Hamiltonian is modified as
\begin{equation}
  \begin{split}
    H^{\boldsymbol{y}} &= P_{\boldsymbol{y}} H P_{\boldsymbol{y}} \\
                       &= \sum_{i} (-1)^{y_i} h_i Z_i + \sum_{i<j} (-1)^{y_i + y_j} J_{i,j} Z_i Z_j.
  \end{split}
  \label{eq:tran_H}
\end{equation}
Although the change of the basis by $P_{\boldsymbol{y}}$ modifies the sign of the coefficients, the eigenvalues of the cost Hamiltonian in the original basis are preserved in the new basis.
We refer to this transformation as a gauge transformation.
The change of basis (bit-flip transformation) also transforms the cost unitary circuit as $ {U_P(\gamma) \to U_P^{\boldsymbol{y}}(\gamma) = \exp(-i\gamma P_{\boldsymbol{y}} H_CP_{\boldsymbol{y}})} $.
However, the overall statistical distribution of measurement outcomes remains unchanged.
We can easily prove this result using $P_{\boldsymbol{y}} U_M P_{\boldsymbol{y}} = U_M$ and $P_{\boldsymbol{y}} \ket{+}^{\otimes n} = \ket{+}^{\otimes n}$.
Details of the proof can be found in the original paper~\cite{Filip2024}.

In the presence of noise, bit-flip gauge invariance no longer holds because the noise can break this bit-flip symmetry.
It implies that different choices of the gauge-transformed cost Hamiltonian \( H^{\boldsymbol{y}} \) lead to different noise effects on the computation, although each $H^{\boldsymbol{y}}$ represents the same problem.
In the NDAR method, we assume the existence of a noise attractor state $\ket{\bm{x}_\mathrm{att}}$, which is a certain computational basis state sampled more frequently than others.
Then, we map the best-found solution $\tilde{\bm{y}}$ to this noise attractor state, and we can construct $H^{\tilde{\bm{y}}}$ based on this mapping.

Identifying a system's noise attractor states under different types of noise presents significant challenges.
However, under amplitude damping noise, low-Hamming-weight states intuitively become the dominant noise attractor states.
Indeed, previous research has demonstrated that $\ket{\bm{x}_\mathrm{att}} = \ket{0\cdots 0}$ serves as a good approximation for the noise attractor state~\cite{Filip2024}.
In this study, we also assume $\ket{\bm{x}_\mathrm{att}} = \ket{0\cdots 0}$ to leverage amplitude damping noise.
With this assumption, we can establish the relationship
\begin{equation}
  \Braket{\tilde{\bm{y}} | H | \tilde{\bm{y}}} \equiv
  \Braket{0 \cdots 0 | H^{\tilde{\bm{y}}} | 0 \cdots 0}.
  \label{eq:H_equality}
\end{equation}
If \( \tilde{\bm{y}} \) corresponds to the best-found solution in the original Hamiltonian, the all-zeros state becomes the optimal energy state in the gauge-transformed Hamiltonian.
The distribution after gauge transformation tends to concentrate around the best-found solution $\tilde{\bm{y}}$.
Thus, in each NDAR iteration, the solution is remapped to the best energy state found in the previous step, followed by sampling under the transformed Hamiltonian.
Consequently, the energy of the all-zeros state gradually improves with each iteration, approaching the global optimal energy.

\section{NDAR with Delay-gate-induced Amplitude Damping}
\label{sec:proposed_method}
\begin{algorithm}[t]
  \caption{Noise-Directed Adaptive Remapping with Delay-Gate-Induced Amplitude Damping (Attractor state $:= |0 \cdots 0\rangle$)}
  \begin{algorithmic}[1]
    \Require
    \Statex $H_0$: Original Hamiltonian for the cost function
    \Statex $U$: Quantum Circuit
    \Statex $T_{\mathrm{d}}$: Duration for delay gate applied before measurement
    \Statex $M$: Number of measurement shots per iteration

    \State Initialize $j \gets 1$
    \While{termination condition not met}
      \State Sample $M$ bitstrings $\{ \boldsymbol{y}^i_j \}_{i=1}^{M}$ using $U$ with $T_{\mathrm{d}}$ delay gate.
      \State Compute energies: $E^i_j = \langle \boldsymbol{y}^i_j | H_{j-1} | \boldsymbol{y}^i_j \rangle$ for all $i$
      \State Select best bitstring $\boldsymbol{y}_{\text{Best}} = \arg\min_i E^i_j$
      \State Apply bit-flip gauge transformation: $H_j := P_{\boldsymbol{y}_{\text{Best}}} H_{j-1} P_{\boldsymbol{y}_{\text{Best}}}$
      \State Increment $j \gets j + 1$
    \EndWhile

  \end{algorithmic}
  \label{alg:ndar}
\end{algorithm}

As described in the previous section, amplitude damping directly influences NDAR performance.
When subjected to stronger amplitude damping noise, quantum states tend to evolve toward low Hamming weight states.
We exploit this property by exploring solutions in the vicinity of the attractor state through quantum circuit sampling.
Subsequently, we perform gauge transformations using the best solution obtained from this sampling process.

This iterative NDAR process parallels classical local search algorithms~\cite{Gendreau2018-jm, Blum2003, Fister2013}.
In best-first local search, each iteration comprises two critical operations: constructing a neighborhood list and identifying the optimal solution within that neighborhood.
The neighborhood typically consists of solutions within a specified Hamming distance from the current solution.
Due to computational efficiency considerations, neighborhood lists often contain only solutions with a Hamming distance of one from the current solution,
although local search algorithm performance depends significantly on the quality of this neighborhood list.
The algorithm explores the best solution within this neighborhood and transitions to the solution.
Through repeated application of these operations, the algorithm progressively discovers higher-quality solutions.
NDAR can be conceptualized as a local search algorithm that generates its neighborhood through quantum circuit sampling and transitions between solutions via gauge transformations.
As sampling outcomes vary with amplitude damping noise strength, the neighborhood characteristics also change accordingly.
Consequently, NDAR's performance can be controlled by modulating the strength of amplitude damping noise.

Based on this observation, we propose a strategy called Delay-Gate-Induced Amplitude Damping, which enhances amplitude damping effects by inserting a delay gate with specified duration $T_{\mathrm{d}}$ before measurement.
Algorithm~\ref{alg:ndar} demonstrates our proposed method.
The NDAR algorithm commences with an initial Hamiltonian \( H_0 \) and systematically samples bitstrings from a quantum circuit $U$ augmented with a delay gate.
At each iteration \( j \), the circuit generates \( M \) bitstrings \( \{ \boldsymbol{y}_j^i \}_{i=1}^M \), which undergo evaluation against the current Hamiltonian \( H_{j-1} \) to yield energies \( E_j^i = \Braket{\boldsymbol{y}_j^i | H_{j-1} | \boldsymbol{y}_j^i} \).
The optimal bitstring \( \boldsymbol{y}_{\text{Best}} = \arg\min_i E_j^i \) is identified based on the problem's objective value.
Upon meeting termination criteria, the algorithm returns the most favorable solution discovered;
otherwise, it applies a bit-flip gauge transformation \( H_j := P_{\boldsymbol{y}_{\text{Best}}} H_{j-1} P_{\boldsymbol{y}_{\text{Best}}} \) to the Hamiltonian.
This transformation effectively reorients the optimization landscape around the attractor state \( |0 \cdots 0\rangle \), facilitating exploration in subsequent iterations with the transformed Hamiltonian.

Previous studies utilized QAOA for sampling within NDAR.
In our work, we also implement sampling with single-layer QAOA.
It has been reported that QAOA parameters exhibit parameter concentration, where similar optimal parameters emerge across various problem instances~\cite{akshay2021parameter,Brandao:2018qoa,Farhi:2019xsx}.
We assume that parameter concentration persists throughout the NDAR iterations since the Hamiltonian structure is preserved during the process.
Consequently, we used the same parameters from the beginning to the end of the iterations.

Our method introduces extended delay times $T_\mathrm{d}$ to enhance the effects of amplitude damping noise.
Quantum decoherence is characterized by two fundamental timescales: thermal relaxation time $T_1$ and dephasing time $T_2$~\cite{Nielsen_Chuang_2010}.
When delay times exceed both timescales, quantum information becomes irretrievably lost to noise.
Thus, we must calibrate the delay time to leverage amplitude damping noise effectively.
From another perspective, NDAR can be seen as a local search algorithm. In this view we can improve solution quality by searching (sampling) the vicinity of the attractor, say known best, state. This idea still remains even if we do not consider the problem information at the sampling process, since the energy is taken into account when gauge transforming the Hamiltonian.
This discussion raises an intriguing question: Can a random circuit combined with NDAR produce results comparable to QAOA?
In the next section we examine QAOA (which encodes the cost Hamiltonian) and random circuits as distinct sampling strategies to investigate this possibility.

\section{Results on the real device}
\label{sec:result}

\subsection{Experimental Setting}
This section presents the experimental results obtained using IBM's Heron quantum processor.
We begin by describing the MaxCut problem addressed in our experiments.
The MaxCut problem represents a fundamental combinatorial optimization problem in graph theory.
Given a graph \( G = (V, E) \) with weighted or unweighted edges, the objective is to partition the node set \( V \) into two disjoint subsets such that the total weight of edges spanning between these subsets is maximized.
The MaxCut problem is formulated as follows:
\begin{equation}
  \max \quad \frac{1}{2} \sum_{(i,j) \in E} w_{i,j}(1 - s_i s_j),
\end{equation}
where $w_{i,j}$ denotes the weight of edge $(i,j)$.
We generated problem instances using graphs characterized by their edge density:
\begin{equation}
  d_{\mathrm{edges}} = \frac{\left| E \right|}{{}_{|V|} \mathrm{C}_2}.
\end{equation}

We evaluate NDAR performance using two distinct problem models.
First, we examine an unweighted graph with $w_{i,j}= 1$ and $d_{\text{edge}}\approx 0.3$.
Second, we investigate a weighted graph with $w_{i,j}=\pm 1$ and $d_{\text{edge}}= 1$, where the both weight signs are taken with equal probability.
In both cases, we address randomly generated instances with $80$ nodes.

Our experiments are conducted on IBM's Heron processor (the $133$-qubit ibm\_torino system).
The median relaxation times are $T_1 \approx 180 \mu$s and $T_2 \approx 127 \mu$s.
We configure delay times at $0$, $50$, and $100~\mu$s, all below the median relaxation timescales.
The QAOA circuits are implemented with a single layer ($p=1$), while the random circuits depth is set to $2$.
Random circuits are generated by applying randomly selected gates from Qiskit's standard gate set to randomly chosen qubits.
Variational parameters are optimized for the original Hamiltonian at iteration $0$ using the noiseless Matrix Product State (MPS) simulator in Qiskit~\cite{qiskit2024}~\footnote{
  Note that the MPS simulator incorporates a different form of approximation due to the truncated bond dimension,
  resulting in simulated sampling that includes these truncation effects.
}, with the maximum bond dimension set to $20$.
For the unweighted MaxCut problem, we obtained optimal parameters $\gamma = -0.152$ and $\beta = 2.041$, while for the fully connected weighted MaxCut problem, the values are $\gamma = -0.171$ and $\beta = 0.905$.
We take $M=1000$ samples at each NDAR iteration for both circuit types.
For each MaxCut problem, we average data from $10$ independent runs per circuit and delay time configuration. 
Maximum number of iteration $n_\mathrm{iter}$ was set to 8 for the unweighted-sparse case and 12 for weighted-dense case, based on empirical testing using Qiskit's fake backend.
We also implemented simulated annealing as a performance baseline using OpenJij~\cite{OpenJij}.

For performance evaluation, we utilize the best energy \( E_{\mathrm{Best}} \) as our primary metric, representing the maximum energy observed in each iteration.

\begin{figure}[t]
  \centering
  \includegraphics[width=0.5\textwidth]{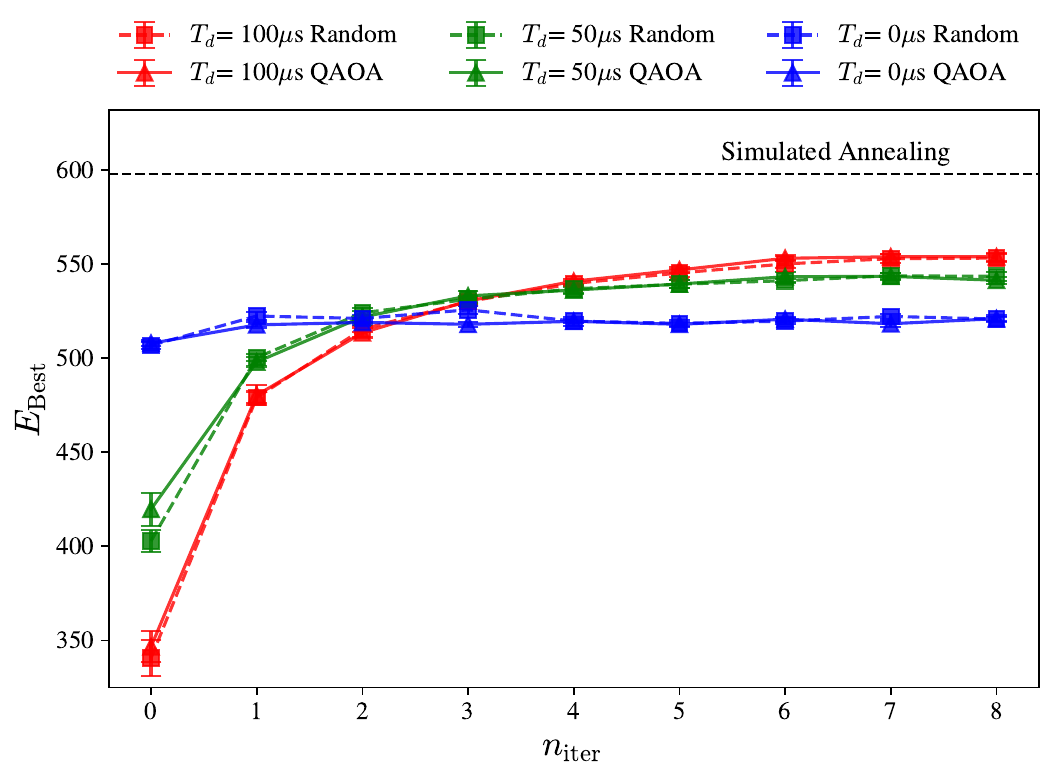}
  \caption{Trajectories of the best energy $E_\mathrm{Best}$ for the unweighted MaxCut problem with edge density $d_{\text{edge}} \approx 0.3$.
    Results are shown for three delay times ($T_{\mathrm{d}}$= 0, 50, 100~$\mu$s) and two circuit types: QAOA and random circuit.}
    \label{fig:80_0.3}
\end{figure}

\begin{figure}[tb]
  \centering
  \includegraphics[width=0.5\textwidth]{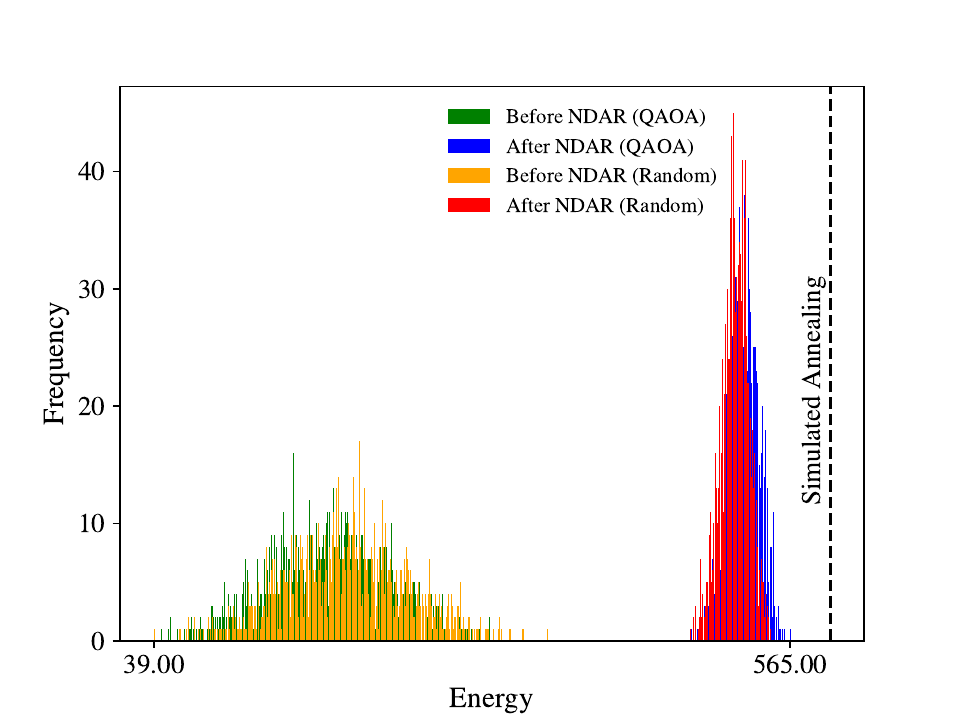}
  \caption{The cost distribution of observed states of unweighted MaxCut problem with a $100\mu$s delay gate, shown for one of the ten runs.
    The dashed line indicates the best energy obtained by the simulated annealing algorithm.
  }
  \label{fig:dist_0.3_100}
\end{figure}

\begin{figure}[tb]
  \centering
  \includegraphics[width=0.5\textwidth]{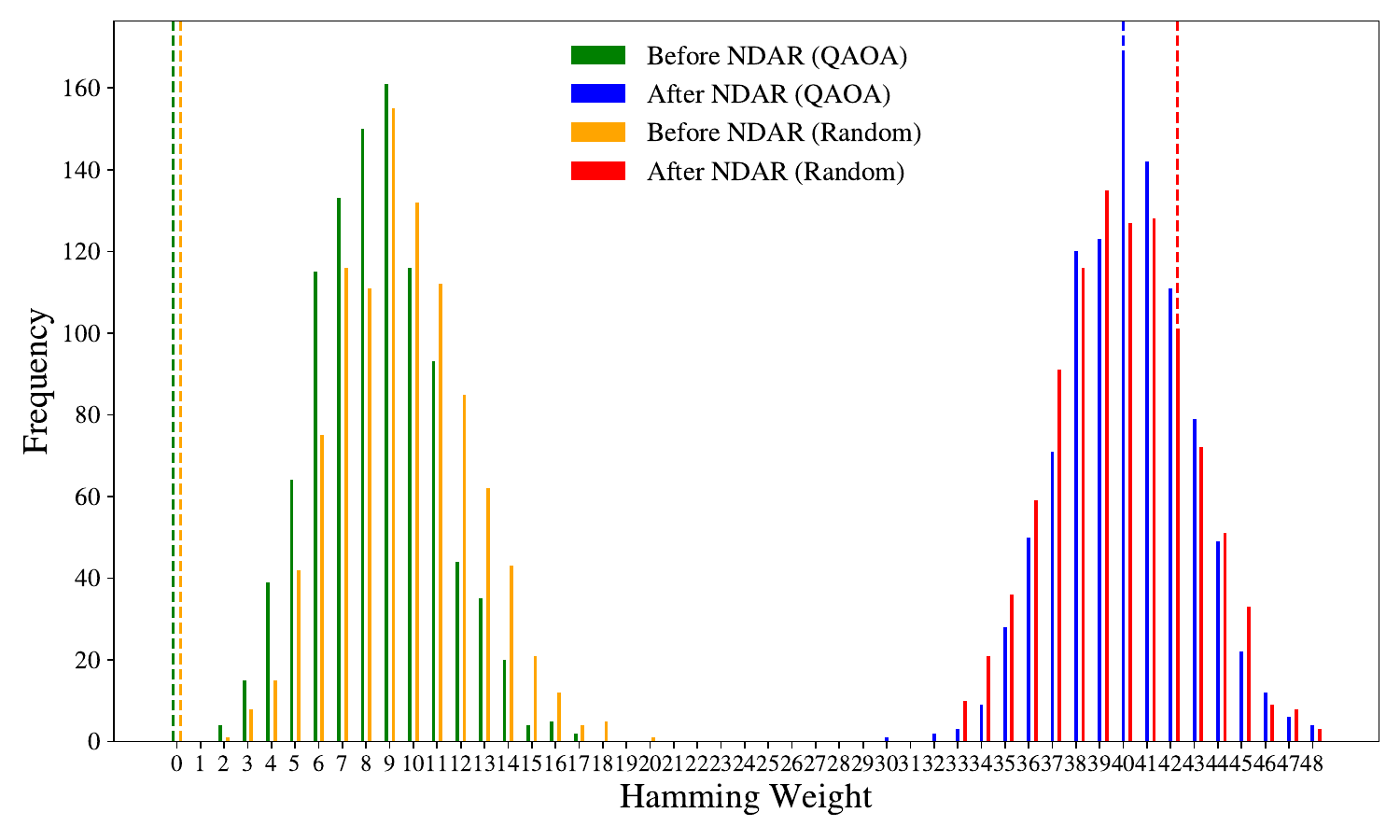}
  \caption{Hamming weights distribution for unweighted MaxCut problem with 100$\mu$s delay time, shown for one of the ten runs.
    Each dashed line indicates the attractor state associated with its corresponding distribution.}
  \label{fig:hw_dist_0.3}
\end{figure}

\begin{figure}[tb]
  \centering
  \includegraphics[width=0.5\textwidth]{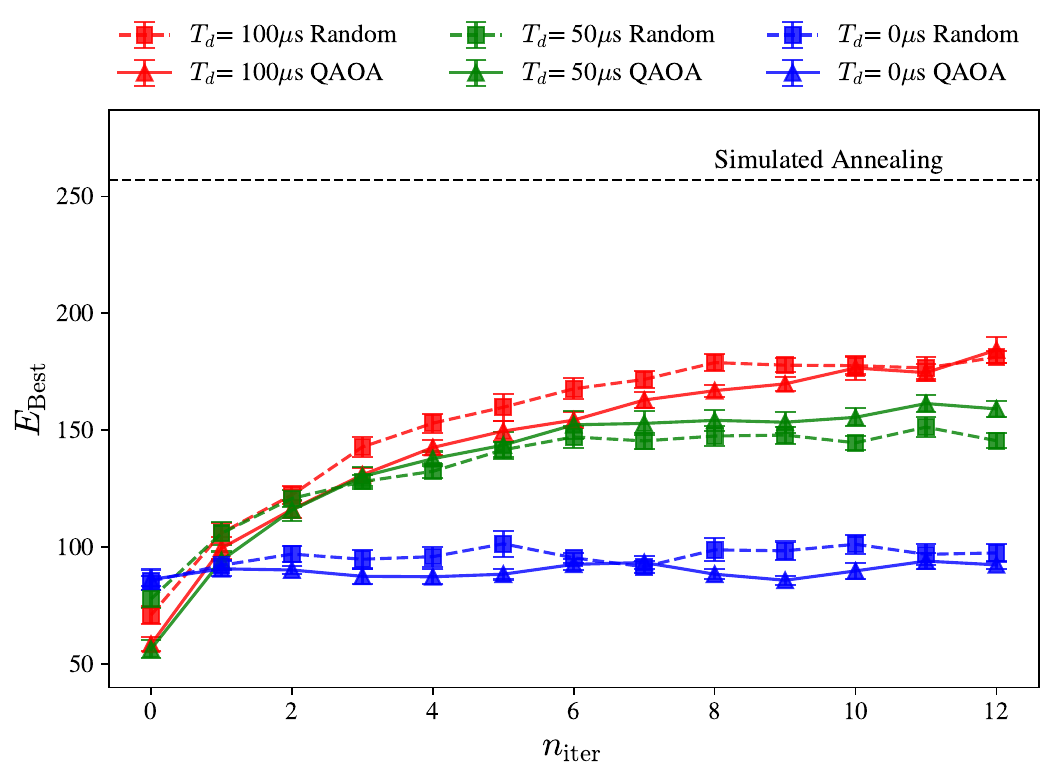}
  \caption{Trajectories of the best energy $E_\mathrm{Best}$ for the weighted MaxCut problem with edge density $d_{\text{edge}}=1$.
    Results are shown for three delay times ($T_{\mathrm{d}}$= 0, 50, 100~$\mu$s) and two circuit types: QAOA and random circuit.}
    \label{fig:80_full}
\end{figure}

\begin{figure*}[t]
  \centering
  \includegraphics[width=1\textwidth]{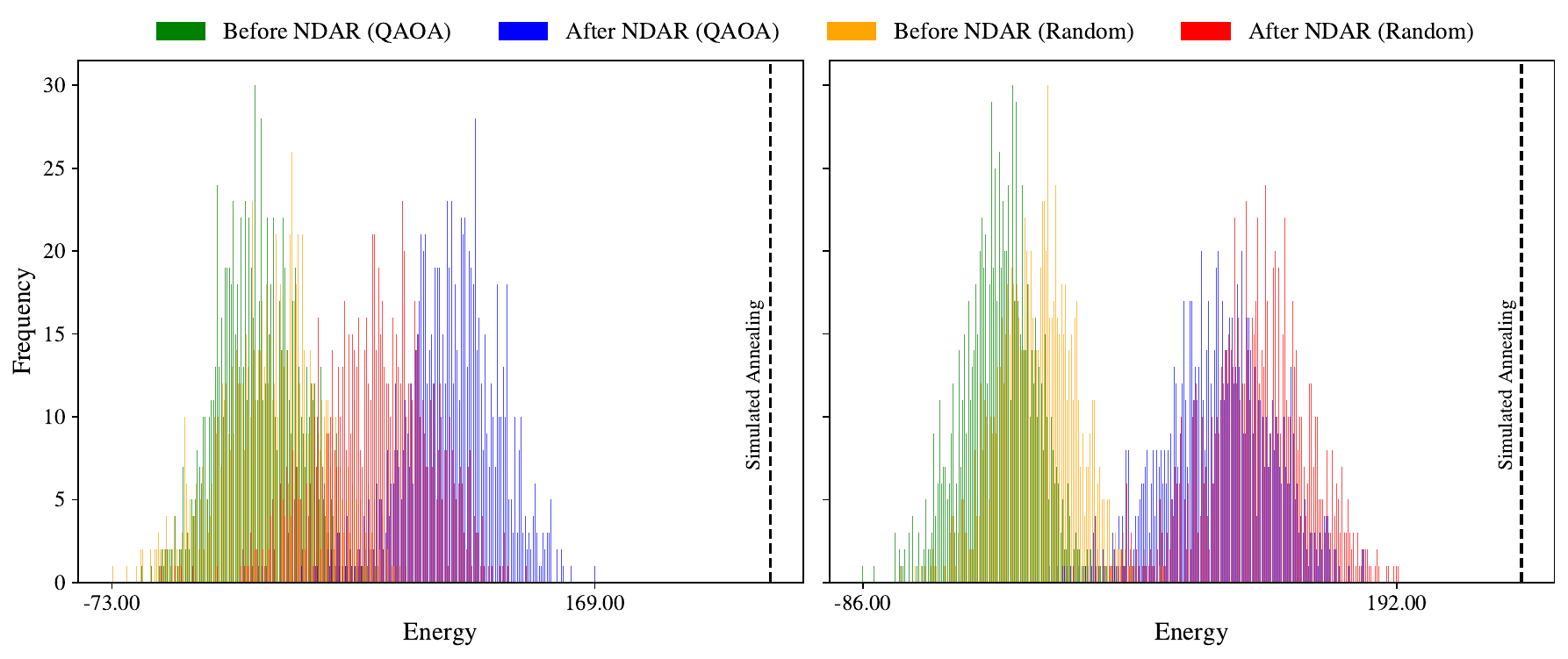}
  \caption{The cost distribution of observed states of a same fully connected weighted MaxCut instance with a $50\mu$s (left) and $100\mu$s (right) delay gate, shown for one of the ten runs.
    The dashed line indicates the best energy obtained by the simulated annealing algorithm.}
    \label{fig:dist_1}

  \centering
  \includegraphics[width=1\textwidth]{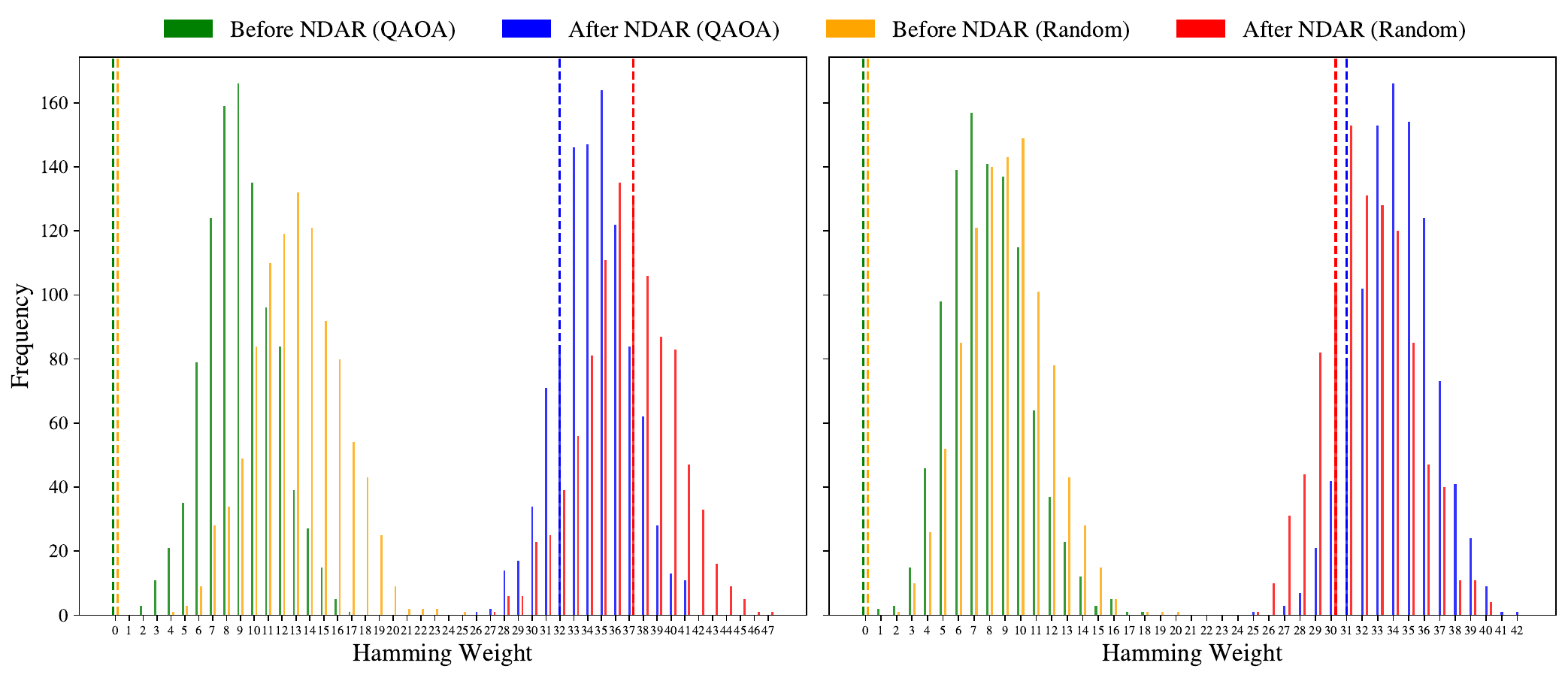}
  \caption{Hamming weights distribution for fully connected weighted MaxCut problem with a $50~\mu$s (left) and $100~\mu$s (right) delay gate, shown for one of the ten runs.
    Each dashed line indicates the attractor state associated with its corresponding distribution.}
    \label{fig:hw_dist_1}
\end{figure*}

\subsection{Experimental Results}

We first examine the low-density ($d_{\text{edge}}\approx 0.3$) unweighted MaxCut problem.
Figure~\ref{fig:80_0.3} displays the best energy $E_\mathrm{Best}$ at each iteration for $T_{\mathrm{d}} = 0,\ 50,\ 100\,\mu$s.
Symbol shapes distinguish circuit types, while colors correspond to different $T_{\mathrm{d}}$ values.
QAOA and random circuits exhibit similar performance patterns across all delay time configurations.
Without delay gates, NDAR fails to demonstrate improvement as iterations progress.
In contrast, NDAR with $T_{\mathrm{d}}=100~\mu$s delay gate produces the lowest $E_\mathrm{Best}$ values at the initial step, but shows consistent improvement with each iteration, achieving the highest $E_\mathrm{Best}$ among all configurations by the final step.
With $T_{\mathrm{d}}=50\mu$s and $100\mu$s settings, the ratio $E_\mathrm{Best} / E_\mathrm{SA}$ reaches 0.908(3) and 0.926(3), respectively.
These results confirm that incorporating delay gates to enhance amplitude damping noise effectively improves NDAR performance.
However, increased amplitude damping also disrupts information within the circuit, and in this experiment, no significant performance difference emerged between QAOA and random circuits.

Notably, QAOA and random circuits exhibit comparable performance even at the initial ($0$-th) iteration, prior to any NDAR transformations.
This similarity stems from hardware noise effects;
under ideal noiseless conditions, QAOA significantly outperforms random circuits, as demonstrated by simulation results where $E_\mathrm{Best}/E_{\mathrm{SA}}$ reaches $0.852(3)$ for QAOA compared to $0.422(47)$ for random circuits.

To quantify the effects of NDAR iterations, we analyzed the cost distributions before and after applying the method with a $T_{\mathrm{d}}=100 \mu$s delay gate, as illustrated in Fig.~\ref{fig:dist_0.3_100}.
The cost distributions for both circuits demonstrate a marked rightward shift, becoming more concentrated around solutions with superior objective values.
This redistribution indicates that the algorithm successfully redirects the attractor state toward regions of the solution space with higher quality solutions.

Next, we examine Hamming weight distributions before and after the NDAR application.
As explained in Sec.~\ref{sec:method}, the Hamiltonian undergoes gauge transformation in each iteration using the best solution \( \tilde{\bm{y}} \) from the previous step.
Figure~\ref{fig:hw_dist_0.3} shows the resulting Hamming weight distributions.
Prior to NDAR iteration, both circuit distributions peak significantly away from the attractor state.
In each iteration, the gauge transformation is applied based on the best solution found, so that the attractor aligns with the optimal energy state from the previous step. Consequently, the attractor's energy gradually improves with each iteration, approaching the global maximum. Furthermore, the attractor state migrates to a Hamming weight of approximately $n/2=40$, which is close to the optimal state typically found in random unweighted MaxCut problems. Over successive NDAR iterations, the distribution peak and the attractor state become closely aligned.

The data presented in Figs.~\ref{fig:dist_0.3_100} and ~\ref{fig:hw_dist_0.3} demonstrate that solutions closer to the attractor (in Hamming distance) 
consistently yield higher objective values.
These findings support the exploitation hypothesis that high-quality solutions exist in the vicinity of other high-quality solutions.

Next, we examine the fully connected ($d_{\text{edge}} = 1$) weighted MaxCut problem.
Figure~\ref{fig:80_full} shows $E_\mathrm{Best}$ trajectories across NDAR iterations.
Definition of symbol/line styles and colors are the same as in previous Fig.~\ref{fig:80_0.3}.
The final solution quality is lower than that of the unweighted-sparse case in Fig.~\ref{fig:80_0.3};
in the current case, QAOA achieves $E_\mathrm{Best} / E_\mathrm{SA}$ ratios of 0.717(21) with $T_{\mathrm{d}}=100\mu$s and 0.619(13) with $T_{\mathrm{d}}=50\mu$s.
Also, the dependence on the circuit type is different from the Fig.~\ref{fig:80_0.3}, particularly at $T_{\mathrm{d}}=50\mu$s, where the QAOA circuit slightly outperforms the random circuit.
This slight discrepancy in performance suggests that the choice of exploration strategy may improve the results.

To elucidate the performance difference at $T_{\mathrm{d}}=50\mu$s delay time, we analyze energy distributions before and after the NDAR iterations at both $T_{\mathrm{d}}=50\mu$s and $T_{\mathrm{d}}=100\mu$s, as shown in Fig.~\ref{fig:dist_1}.
These distributions represent a representative sample selected from ten independent experimental runs.
For the $T_{\mathrm{d}}=100\mu$s configuration, we observe behaviors similar to the unweighted MaxCut case, with both circuit types producing energy distributions that shift toward higher values.
At $T_{\mathrm{d}}=50\mu$s, however, the QAOA circuit exhibits a more pronounced rightward shift in energy distribution than the random circuit.
These results imply that QAOA can provide a more sophisticated exploration strategy compared to random sampling, although this observation is limited to the weighted-dense case with $T_{\mathrm{d}} = 50~\mu\mathrm{s}$.

We further analyze Hamming weight distributions before and after NDAR application at $T_{\mathrm{d}}=50\mu$s and $T_{\mathrm{d}}=100\mu$s delay times, shown in Fig.~\ref{fig:hw_dist_1}.
With $T_{\mathrm{d}}=100\mu$s, both circuits achieve high performance.
After NDAR iteration, we observe that the attractor states exhibit lower Hamming weights than the distribution peaks.
With $T_{\mathrm{d}} = 50~\mu\mathrm{s}$, the QAOA behavior is similar to the $T_{\mathrm{d}} = 100~\mu\mathrm{s}$ case, where the attractor state lies far from the peak of the distribution. In contrast, for the random circuit, the positions of the distribution peak and attractor state remain close to each other.
These results suggest that solutions are trapped near the attractor state with insufficient exploration capability to discover higher-quality solutions.
Despite the similar widths of Hamming weight distributions between both circuits, the higher performance of QAOA suggests the importance of incorporating energy information in sampling---an inherent advantage of QAOA as it encodes the cost Hamiltonian.

Note that, in the previous work by Maciejewski \textit{et al.}~\cite{Filip2024}, the NDAR with single-layer QAOA was quite successful with recording higher approximation ratios compared to our current work,
although their target graph was fully connected and also other experimental settings were similar to ours.
Three key methodological differences distinguish our approach from their study:
first, their hardware was quite different from our environment;
second, we utilized delay gates to enhance the amplitude damping noise;
third, we employed a different strategy for QAOA parameter optimization.
The third, algorithmic difference relates to parameter transferability on real devices.
In their study, QAOA parameters were re-optimized at each NDAR step, whereas our approach is based on parameter transferring, eliminating the need for re-optimization at each iteration.
Although we observed QAOA slightly outperforms random circuits at $T_{\mathrm{d}}=50\mu$s, there is room to consider that the parameter transferability can be strikingly harmed under amplitude damping or other types of hardware noise.

\section{Classical NDAR}
\label{sec:classical}

\begin{algorithm}[t]
  \caption{Classical Noise-Directed Adaptive Remapping}
  \begin{algorithmic}[1]
    \Require
    \Statex $C_0$: Original cost function
    \Statex $q$: Bit-suppress probability
    \Statex $n_\mathrm{shots}$: Number of sampling shots per iteration

    \State Initialize $j \gets 1$
    \While{termination condition not met}
      \State Sample $M$ bitstrings $\{ \boldsymbol{y}^i_j \}_{i=1}^{M}$ with suppressing each bit to 0 with probability $q$
      \State Compute energies: $E^i_j = C_{j-1}(\boldsymbol{y}^i_j)$ for all $i$
      \State Select best bitstring $\boldsymbol{y}_{\text{Best}} = \arg\min_i E^i_j$
      \State Apply bit-flip transformation: $C_j := C^{\boldsymbol{y}_{\text{Best}}}_{j-1}$
      \State Increment $j \gets j + 1$
    \EndWhile
  \end{algorithmic}
  \label{alg:cls_ndar}
\end{algorithm}

\subsection{Details of Classical NDAR}
In the previous section, we implemented the NDAR method on a quantum device to solve an $80$-variable optimization problem.
Experiments on even bigger systems are restricted by current quantum hardware limitations.
To address this constraint and to make our local-search fashion perspective more concrete, we propose a classical algorithm inspired by the NDAR method as a simplified approach to validate the framework's effectiveness for larger problems.
Our classical implementation emulates NDAR behavior with a random circuit by sampling solutions near the attractor state and iteratively updating this state to discover progressively higher-quality solutions.
Unlike QAOA-based NDAR, which incorporates energy information through Hamiltonian encoding, our classical sampling does not consider solution energies during the sampling process.
Consequently, this classical implementation serves primarily as a performance baseline rather than demonstrating the classical simulatability of the quantum NDAR method.
We refer to this approach as ``classical NDAR method''.

Algorithm~\ref{alg:cls_ndar} outlines the classical NDAR's workflow.
The algorithm begins with bitstring sampling to generate candidate solutions.
To emulate quantum NDAR behavior, we generate $n_\mathrm{shots}$ bitstrings, with each bit independently set to $0$ with probability $q$ and $1$ with probability $1 - q$.
Higher $q$ values increase the likelihood of sampling solutions with smaller Hamming distances from the attractor state.
We evaluate each bitstring according to the cost function in Eq.~\eqref{eq:isingcost}, where bits are converted to spins via $s_i = 1 - 2x_i$.
After evaluation, the bitstring with the best objective value is selected as the new attractor state.
A classical bit-flip transformation is then applied to the cost function using this selected bitstring.
Instead of using the quantum bit-flip operator $P_{\boldsymbol{y}}$, we transform each term in Eq.~\eqref{eq:isingcost} according to the pattern shown in Eq.~\eqref{eq:tran_H}:
\begin{equation}
  \begin{split}
    C^{\boldsymbol{y}}(\boldsymbol{s}) &= \sum_{i} (-1)^{y_i} h_i s_i + \sum_{i<j} (-1)^{y_i +y_j}J_{ij} s_i s_j.
  \end{split}
  \label{eq:updated_cost}
\end{equation}
This iterative process continues with the updated cost function until termination criteria is satisfied.

We evaluated the classical NDAR performance through $10$ independent runs for each probability value $q \in \{0.85, 0.90, 0.95\}$.
We evaluated the performance using $80$-node and $300$-node unweighted and weighted MaxCut instances.
For the $80$-node problems, we used the same instances examined in the previous section. The maximum number of iterations $n_\mathrm{iter}$ was set to 12 and 15 for the unweighted and weighted MaxCut problems on 80-node instances, respectively, and to 20 and 25 for the corresponding 300-node problems, following a similar configuration.

\subsection{Experimental Results}

\begin{figure}[tb]
  \centering
  \includegraphics[width=0.5\textwidth]{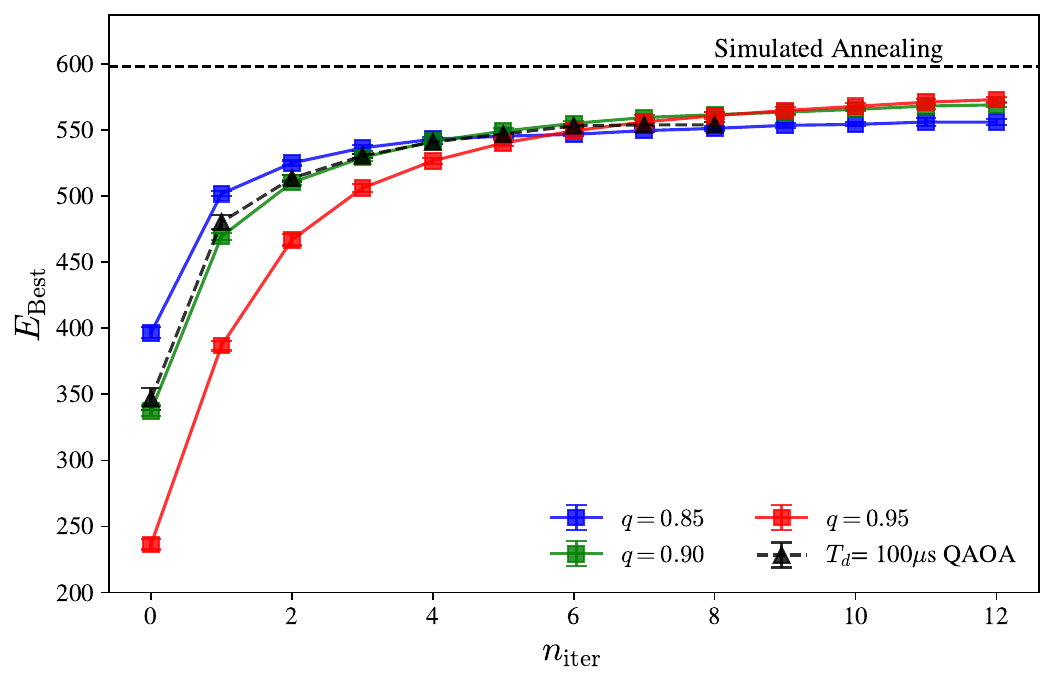}
  \caption{Trajectories of the best energy $E_\mathrm{Best}$ for the unweighted MaxCut problem with edge density $d_{\text{edge}} \approx 0.3$ using the classical NDAR method.
    Results are shown for three probabilities $q = 0.85, 0.90$ and $0.95$ alongside with $T_{\mathrm{d}}=100\mu$s QAOA.}
    \label{fig:classical_0.3}
\end{figure}

\begin{figure}[tb]
  \centering
  \includegraphics[width=0.5\textwidth]{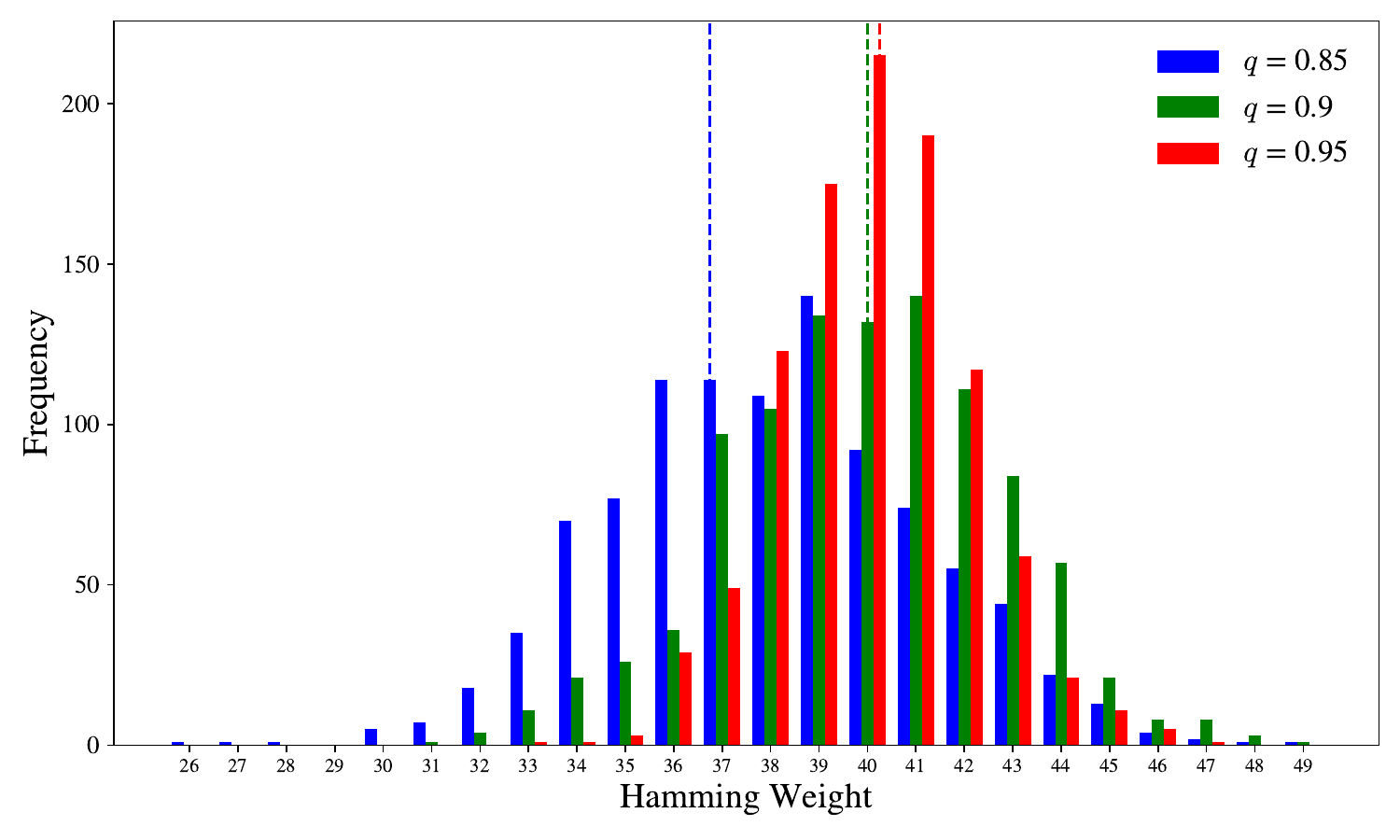}
  \caption{Hamming weights distribution for unweighted MaxCut problem under three different $q$ = 0.85, 0.9, and 0.95.
    The dashed lines refer to their own attractor state.}
  \label{fig:hw_classical_0.3}
\end{figure}

We first examine the low-density ($d_{\text{edge}} \approx 0.3$) 80-node unweighted MaxCut problem.
Figure~\ref{fig:classical_0.3} plots $E_\mathrm{Best}$ trajectories for probability values $q \in \{0.85, 0.90, 0.95\}$.
Each line color corresponds to a different probability value $q$, with results from QAOA-based NDAR at $T_{\mathrm{d}}=100\mu$s included for comparison.
The classical NDAR exhibits performance characteristics similar to those of the quantum NDAR.
Notably, the trajectory for $q=0.9$ shows remarkable similarity to that of QAOA-based NDAR.
Within the classical NDAR framework, increasing $q$ value leads to slower convergence speed but eventually produces better final objective values.
With $q = 0.95$, which yields the best performance, the ratio $E_{\mathrm{Best}} / E_{\mathrm{SA}}$ reaches 0.958(3), exceeding 0.926(3) achieved by QAOA-based NDAR.

Figure~\ref{fig:hw_classical_0.3} illustrates the Hamming weight distributions resulting from the classical NDAR application across different $q$ values.
These distributions represent a typical outcome selected from ten independent trials.
As previously noted, optimal energy states in unweighted MaxCut problems typically exhibit Hamming weights near $n/2$.
Indeed, for both $q = 0.90$ and $q = 0.95$, the attractor states converge to Hamming weight 40.
Similarly, in quantum implementations as shown in Figure~\ref{fig:hw_dist_0.3}, distribution peaks tend to concentrate near their respective attractor states.
This further confirms that solutions with smaller Hamming distances from optimally positioned attractor states consistently produce better objective values.

Next, we evaluate the classical NDAR performance on the fully connected ($d_{\text{edge}}=1$) weighted MaxCut problem.
Figure~\ref{fig:classical_1} shows $E_\mathrm{Best}$ trajectories for probability values $q \in \{0.85, 0.90, 0.95\}$.
The performance pattern resembles that of the unweighted-sparse case, though the maximum $E_{\mathrm{Best}} / E_{\mathrm{SA}}$ ratio decreases to 0.874(22).
Consistent with the unweighted-sparse case, performance improves with increasing $q$ values.
However, the performance differences between various $q$ values are more pronounced than in the unweighted-sparse case.

Figure~\ref{fig:hw_classical_1} displays Hamming weight distributions for the fully connected weighted MaxCut problem after the classical NDAR application.
Similar to the unweighted-sparse case, distribution peaks closely align with their corresponding attractor states.
We can also observe that the $q$-dependence is more evident than in the unweighted-sparse case in terms of the Hamming weight too.
These results collectively demonstrate that the classical NDAR performance improves when sampling more heavily concentrates around the attractor state in Hamming distance space.

\begin{figure}[tb]
  \centering
  \includegraphics[width=0.5\textwidth]{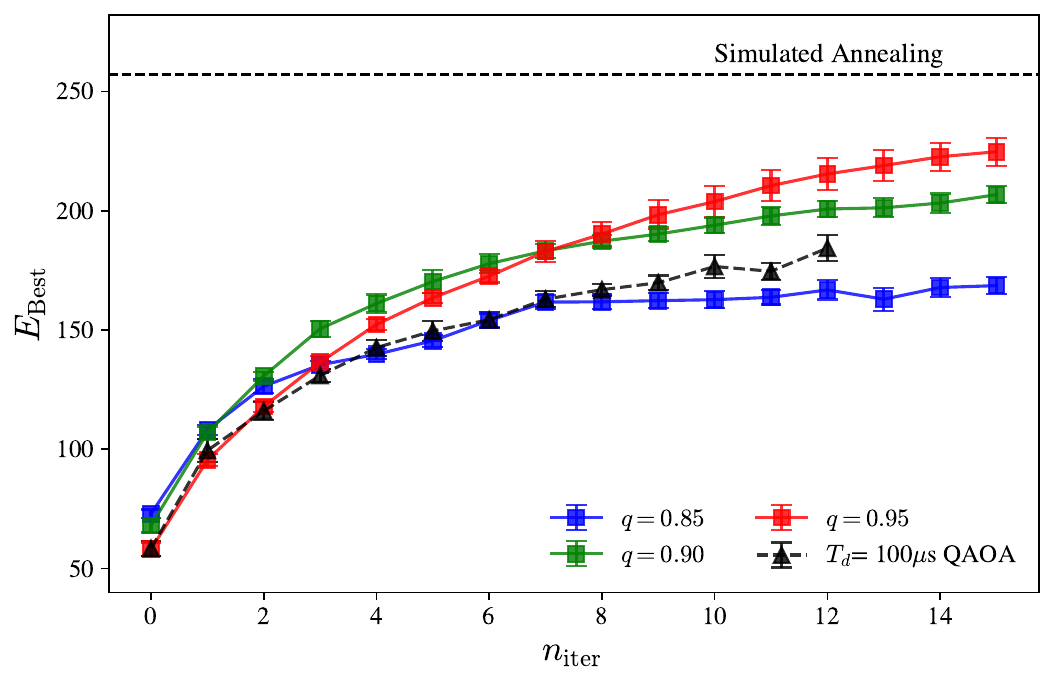}
  \caption{Trajectories of the best energy $E_\mathrm{Best}$ for the weighted MaxCut problem with edge density $d_{\text{edge}} = 1$ using the classical NDAR method.
    Results are shown for three probabilities $q = 0.85, 0.90$ and $0.95$ alongside with $T_{\mathrm{d}}=100\mu$s QAOA.}
    \label{fig:classical_1}
\end{figure}

\begin{figure}[tb]
  \centering
  \includegraphics[width=0.5\textwidth]{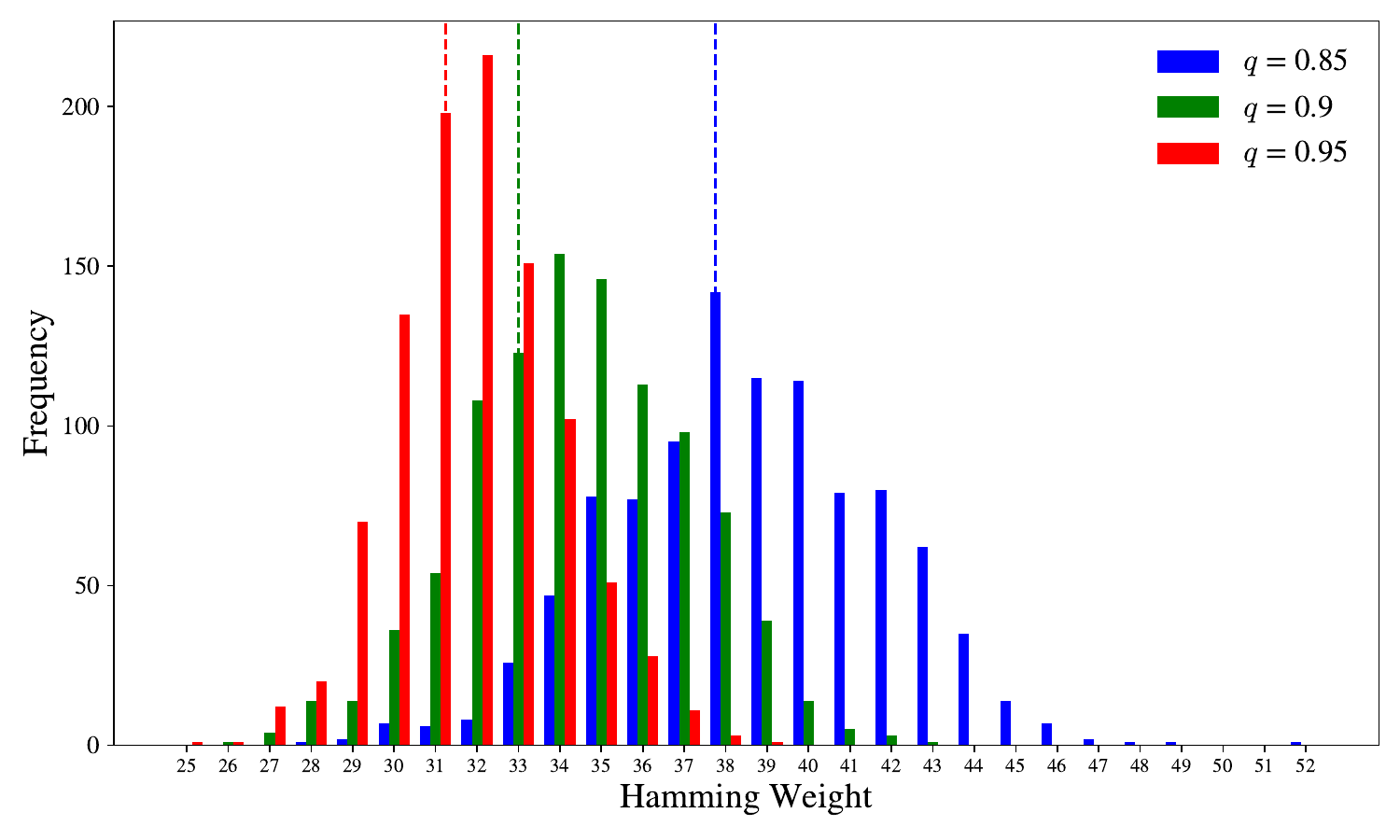}
  \caption{Hamming weights distribution for fully connected weighted MaxCut problem under three different $q$ = 0.85, 0.9, and 0.95.
    The dashed lines refer to their own attractor state.}
  \label{fig:hw_classical_1}
\end{figure}

\begin{figure}[tb]
  \centering
  \includegraphics[width=0.5\textwidth]{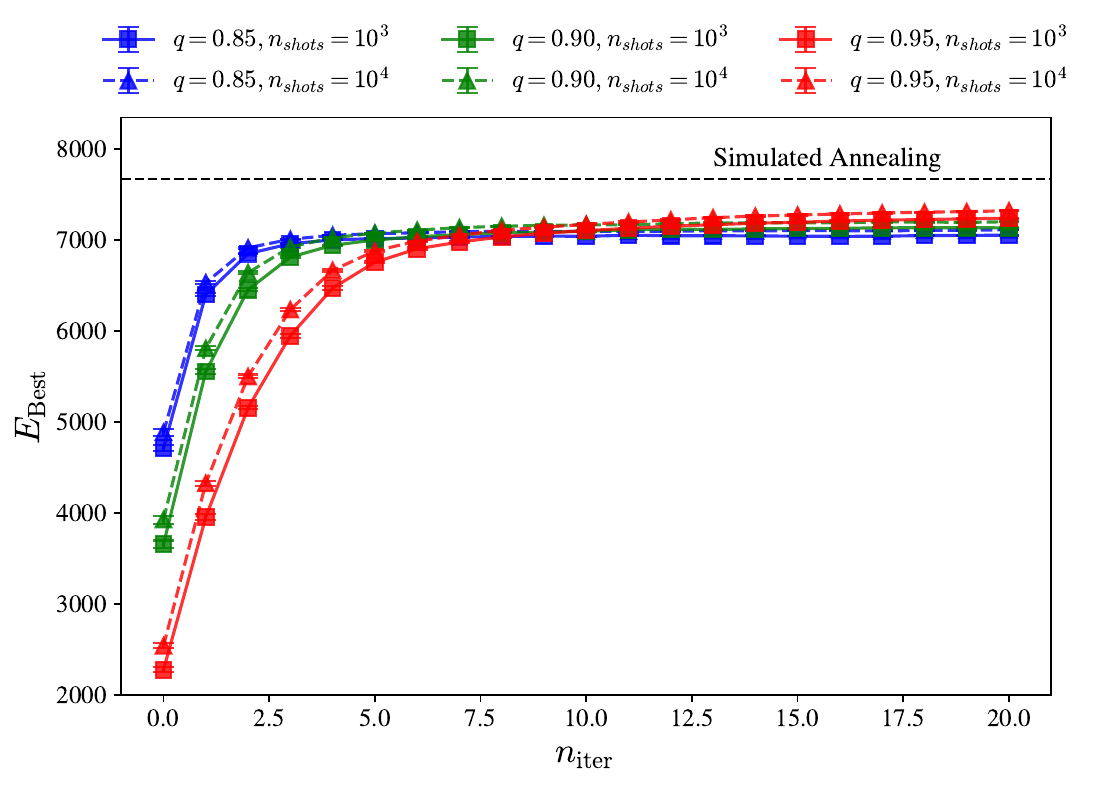}
  \caption{Trajectories of the best energy $E_\mathrm{Best}$ for the weighted MaxCut problem on a 300-node graph with edge density $d_{\text{edge}} \approx 0.3$, using the classical NDAR method. Results are shown for three probability settings, $q = 0.85$, $0.90$, and $0.95$, under both $10^3$ and $10^4$ sampling shots.}
    \label{fig:classical_0.3_300}
\end{figure}

\begin{figure}[tb]
  \centering
  \includegraphics[width=0.5\textwidth]{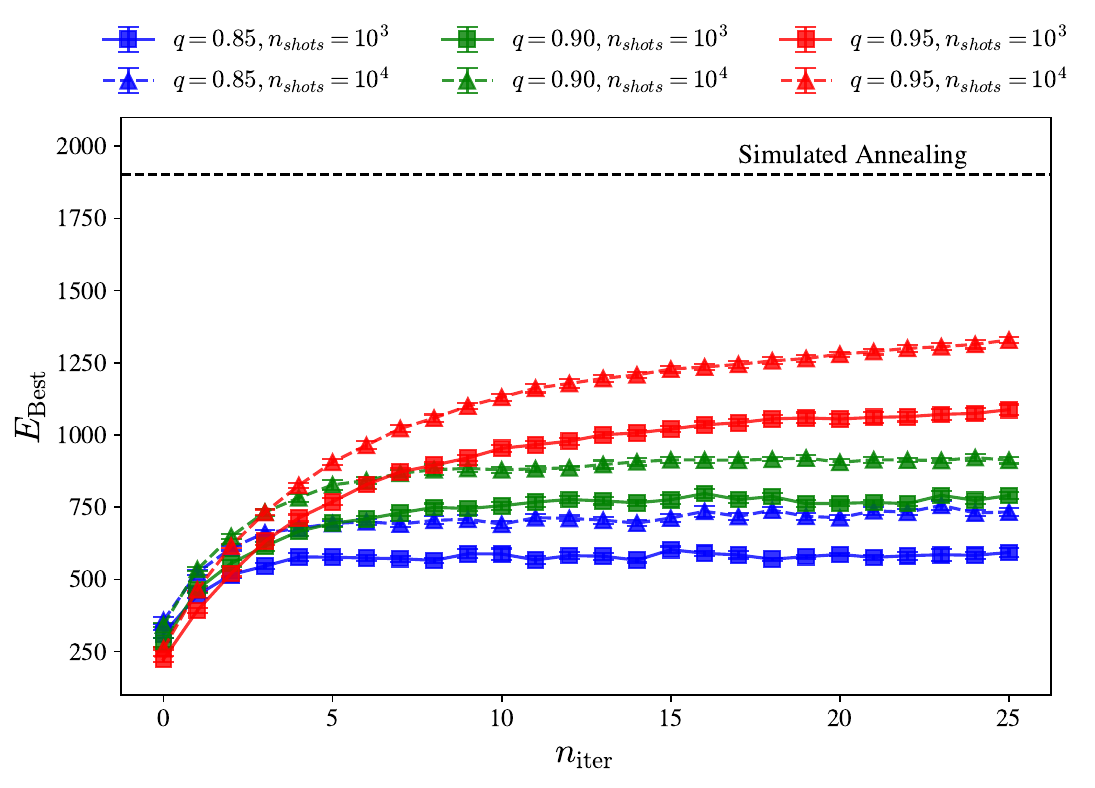}
  \caption{Trajectories of the best energy $E_\mathrm{Best}$ for the weighted MaxCut problem on a 300-node graph with edge density $d_{\text{edge}} = 1$, using the classical NDAR method. Results are shown for three probability settings, $q = 0.85$, $0.90$, and $0.95$, under both $10^3$ and $10^4$ sampling shots.}
    \label{fig:classical_1_300}
\end{figure}

We extend our study to MaxCut problems on $300$-node graphs.
We first examine the low-density unweighted MaxCut case by varying the number of sampling shots $n_\mathrm{shots}$ as shown in Fig.~\ref{fig:classical_0.3_300}.
Similar to the 80-node case, increasing the probability parameter $q$ leads to improved final objective values.
Moreover, increasing $n_\mathrm{shots}$ slightly improves the final objective values.
We obtained $E_{\mathrm{Best}} / E_{\mathrm{SA}} \approx 0.9548(9)$ at $q = 0.95$ and $n_\mathrm{shots} = 10^4$.
We also observed a similar trend in the weighted MaxCut case in Fig.~\ref{fig:classical_1_300}.
The performance differences between various $q$ and $n_\mathrm{shots}$ values are more pronounced in the weighted-dense case than in the unweighted-sparse case.
With $q = 0.95$ and $n_\mathrm{shots} = 10^4$, the final objective value achieves $ E_{\mathrm{Best}} / E_{\mathrm{SA}} \approx 0.698(6)$.
This dependency on  $n_\mathrm{shots}$ can be attributed to the polynomial increase in the number of solutions close to the attractor state in terms of Hamming distance as the size of problem increases.

\section{Summary}
\label{sec:summary}

In this work, we implemented the NDAR method on an IBM's real devices, and presented a classical heuristic perspective on its underlying mechanism.
NDAR alternates between exploration---through sampling solutions from a quantum circuit---and exploitation---by transforming the cost Hamiltonian based on the best solution found.
In classical heuristic algorithm design, balancing exploration and exploitation proves essential for efficient solution space search,
and the NDAR method provides a practical framework for managing this balance through controlled sampling and strategic utilization of noise.
We varied the exploration strategy by testing two different circuits (QAOA and random circuits) and altered the exploitation strength by adjusting the delay time $T_{\mathrm{d}}$.

We evaluated the performance of NDAR on unweighted-sparse and weighted-dense MaxCut problems with 80 nodes, using the best energy $E_{\mathrm{Best}}$ obtained at each iteration as our performance metric.
In the low-density unweighted MaxCut case ($d_{\text{edge}} \approx 0.3$), both circuit types demonstrated strong performance with $T_{\mathrm{d}}=50\mu$s and $T_{\mathrm{d}}=100\mu$s delay settings, achieving final $E_{\mathrm{Best}} / E_{\mathrm{SA}}$ values exceeding 0.9.
For the fully connected weighted MaxCut problem ($d_{\text{edge}} = 1$), the observed performance was analogous to that in the low-density unweighted case, where longer delay times consistently yielded better solutions.
However, the solution quality remained lower than that in the unweighted-sparse case.
In this experiment, QAOA circuits achieved $E_{\mathrm{Best}} / E_{\mathrm{SA}}$ ratios of 0.72 with $T_{\mathrm{d}} = 100\mu$s and 0.62 with $T_{\mathrm{d}} = 50\mu$s.
Notably, under $T_{\mathrm{d}}=50\mu$s delay conditions, QAOA slightly outperformed the random circuit in terms of solution quality.

To address current quantum hardware limitations and to clarify our local-search-inspired perspective, we developed the classical variant of NDAR inspired by our experimental findings.
This classical implementation samples solutions near the attractor state and iteratively refines refines it, mirroring the original NDAR procedure based on quantum sampling.
It emulates random-circuit NDAR without incorporating energy information during sampling, unlike QAOA-based NDAR with its Hamiltonian encoding mechanism.
Consequently, this classical implementation serves as a performance baseline for evaluating NDAR effectiveness.
In this variant, each bit in a bitstring is independently set to $0$ with probability $q$ and to $1$ with probability $1-q$.
Higher $q$ values increase the likelihood of sampling solutions with smaller Hamming distances from the attractor state.
For both MaxCut problem formulations, performance improved as $q$ increases from 0.85 to 0.95.
The performance differential between $q$ values was more pronounced in the fully-connected weighted MaxCut problem compared to the unweighted-sparse case.
With $q = 0.95$, classical NDAR achieved remarkable $E_{\mathrm{Best}} / E_{\mathrm{SA}}$ values of $0.96$ and $0.87$ for the unweighted-sparse and the weighted-dense $80$-node problem.
We also examined the $300$-node problem and demonstrated that the same trend exists.
These baseline results from the classical NDAR suggest that the quantum NDAR can potentially solve larger problems.

Previous research~\cite{Filip2024} has demonstrated that NDAR with optimization of variational parameters at each step performs effectively for fully connected problems.
Reducing computational overhead is a key challenge for practical NDAR implementation.
Integrating NDAR with QAOA parameter transfer offers a promising approach to address this challenge, minimizing computational requirements by eliminating the need for repeated variational parameter optimization.
However, our study did not demonstrate significant performance advantages for parameter-transferred QAOA over random circuits in most settings.
Although parameter transferability has been demonstrated on real quantum devices in several studies~\cite{Montanez-Barrera:2024pax,Montanez-Barrera:2024tos,OLeary:2025mdw}, its effectiveness for large-scale problems under the influence of amplitude damping noise requires further investigation.
Therefore, a critical direction for future research lies in systematically characterizing parameter transferability in noisy quantum environments, particularly for large-scale optimization problems where computational efficiency is paramount.

Another research direction is to investigate how we can prepare or determine an attractor state other than by enhancing the amplitude damping noise.
The current delay-gate-induced approach inevitably affects other types of noise besides amplitude damping noise.
Experiments with the classical NDAR suggest that constructing an attractor state manually (\textit{e.g.}, randomly setting some bits in a bitstring obtained from QAOA to 0) may be the simplest approach, although its performance remains to be investigated.

\section*{Acknowledgment}
We acknowledge the use of IBM Quantum services for this work.
The views expressed are those of the authors and do not reflect the official policy or position of IBM or the IBM Quantum team.

\printbibliography

\end{document}